\documentclass[english]{revtex4-1}
\usepackage[T1]{fontenc}
\usepackage[latin9]{inputenc}
\usepackage{amsmath}
\usepackage{amssymb}
\usepackage{esint}
\usepackage{babel}
\begin{document}

\title{Infrared Divergence Separated for Stochastic Force\\
- Langevin Evolution in the Inflationary Era - }

\author{Masahiro Morikawa}

\email{hiro@phys.ocha.ac.jp}

\affiliation{Department of Physics, Ochanomizu University, 2-1-1 Otsuka, Bunkyo,
Tokyo 112-8610, Japan}
\begin{abstract}
Inflation in the early Universe is a grand phase transition which
have produced the seeds of all the structures we now observe. We focus
on the non-equilibrium aspect of this phase transition especially
the inevitable infrared (IR) divergence associated to the the quantum
and classical fields during the inflation. There is a long history
of research for removing this IR divergence for healthy perturbation
calculations. On the other hand, the same IR divergence is quite relevant
and have developed the primordial density fluctuations in the early
Universe. We develop a unified formalism in which the IR divergence
is clearly separated from the microscopic quantum field theory but
only appear in the statistical classical structure. We derive the
classical Langevin equation for the order parameter within the quantum
field theory through the instability of the de Sitter vacuum during
the inflation. This separation process is relevant in general to develop
macroscopic structures and to derive the basic properties of statistical
mechanics in the quantum field theory. 
\end{abstract}
\maketitle

\section{introduction}

Primordial quantum fluctuations have produced the present macroscopic
structures of the Universe\cite{baumann2009}. Quantum fields are
squeezed in the exponentially expanding de Sitter Universe (i.e. inflation
\cite{sato1981,Guth1981}) and develop the seeds of density perturbations,
which eventually grow into the galaxies and the clusters we now observe.
These seed fluctuations have almost the Zeldovich spectrum which diverges
in the infrared realm. This infrared (IR) divergence is induced by
the massless minimally coupled scalar condensation degrees of freedom
in the de Sitter space. The IR divergence is thus necessary and unavoidable
process for producing macroscopic realm. 

On the other hand, the same IR divergence of the same quantum fields
in de Sitter space destroys the perturbation evaluation of higher
order corrections. There are many references trying to avoid this
catastrophic of the theory\cite{Polyakov2008,arai2012}. Thus the
IR divergence is highly unfavorable in the microscopic realm. 

The above dilemma of IR divergence opposing in micro-macro realms
with each other may be deeply related with the problem of the transition
from quantum fluctuations into classical density perturbations in
the inflationary era, which itself is a grand phase transition. In
this phase transition, the density perturbation is the order parameter,
which violates the translational invariance in 3D real space\cite{Morikawa1987}.
This transition problem has been often discussed in various aspects
such as the classicalization when passing through the horizon, the
frozen fluctuations, squeezing of the vacuum, decoherence, or dis-entanglement,...
However any comprehensive description has not yet given so far \cite{Nambu2009}. 

The above problems would become clear if we adopt that a quantum field
has two phases each represents micro- and macro- degrees of freedom.
The macroscopic classical degrees of freedom is the condensation of
the quantum field $\varphi$ and the microscopic quantum degrees of
freedom is the quantum excitations $\hat{\phi}$ on the classical
filed \cite{Fukuyama2009}. 

The separation of them becomes clear if we use the generalized effective
action method \cite{Morikawa1986}. This method is briefly introduced
in section 2. According to this formalism, the two kinds of degrees
of freedom $\varphi$ and $\hat{\phi}$interact with each other. Furthermore
the IR divergence turns out to appear only in the statistical part
of the Langevin dynamics for $\varphi$. This separation therefore
makes the ordinary perturbation calculations possible in the quantum
field theory for $\hat{\phi}$. 

The popular stochastic method \cite{starobinsky1982} fits well with
this formalism (section 3) although the artificial separation of the
field is necessary. This problem is resolved if we introduce genuine
interaction term (section 4), which clearly defines the classical
order parameter. We further examine the IR property of the general
massive non-minimally coupled scalar field in the de Sitter spacetime
(section 5). Lastly we summarize our work and comment on the general
generation of statistical mechanics (section 6).

\section{langevin equation from quantum field theory}

We will derive Langevin equation in de Sitter space later. In this
section, we start from the classical Langevin equation back to the
field theory. 

Langevin equation is a typical description of a particle motion exerted
by both the potential force $-V'$ and the random force $\xi$ with
friction $\gamma$ : 
\begin{equation}
\ddot{x}(t)=-\gamma\dot{x}(t)-V'(x(t))+\xi(t)\label{eq:Langevinoriginal}
\end{equation}
where the random field obeys the statistical property determined by
the weight functional $P[\xi]$, 
\begin{equation}
\left\langle ...\right\rangle _{\xi}=\int D[\xi]...P[\xi],
\end{equation}
where we temporally assume the Gaussian form $P[\xi]=e^{-\int\xi(t)^{2}/(2\sigma^{2})}$.
Then the $x-$correlation can be generated by 
\begin{eqnarray}
Z[J] & \equiv & \left\langle e^{-i\int dtJ(t)x(t)}\right\rangle _{\xi}\\
 & = & \int D[\xi]D[x]P[\xi]\delta[\ddot{x}(t)+\gamma\dot{x}(t)+V'(x(t))-\xi(t)]e^{-i\int dtJ(t)x(t)}\nonumber \\
 & = & \int D[\xi]D[x]D[x']P[\xi]e^{i\int dtx'(t)\{\ddot{x}(t)+\gamma\dot{x}(t)+V'(x(t))-\xi(t)\}}e^{-i\int J(t)x(t)}\nonumber \\
 & \equiv & \int D[\xi]D[x]D[x']P[\xi]e^{i\tilde{S}[x,x']-i\int Jx}\nonumber 
\end{eqnarray}
where the integral form of the delta functional is utilized, and 
\begin{equation}
\tilde{S}[x,x']\equiv\int dt\{-\dot{x}'(t)\dot{x}(t)+\gamma x'(t)\dot{x}(t)+x'(t)V'(x(t))-x'(t)\xi(t)\}
\end{equation}
where the boundary term is dropped. This is the 'action' because the
application of the least action principle for the variable $x'(t)$
yields the original Langevin equation Eq.(\ref{eq:Langevinoriginal}).
There is another expression for $Z[J]$ given by integrating out $\xi$,
\begin{equation}
Z[J]=\int D[x]D[x']e^{i\tilde{\Gamma}[x,x']-i\int Jx}
\end{equation}
where the 'complex action' is, 
\begin{equation}
\tilde{\Gamma}[x,x']\equiv\int dt\{-\dot{x}'(t)\dot{x}(t)+\gamma x'(t)\dot{x}(t)+x'(t)V'(x(t))-i\sigma^{2}x'(t)^{2}/2\}.
\end{equation}
It is apparent that the imaginary part of $\tilde{\Gamma}[x,x']$
represents statistical fluctuations. 

It is possible to reverse the logic. If we have a complex action including
an extra degrees of freedom like $x'$ above, we can derive a Langevin
equation. This is the formalism of the generalized effective action
method utilizing the closed time-contour \cite{Morikawa1986}. This
formalism is a slight generalization of the ordinary quantum field
theory but particularly suitable for the dissipative dynamics for
the condensed classical variables\cite{keldish1964,kadanoff1962}. 

Let us consider the quantum field theory generalizing the above considerations.
The generating functional of the many point functions is defined as
\begin{eqnarray}
\tilde{Z}[\tilde{J}] & \equiv & {\rm Tr}\left[\tilde{T}\left[\exp[i\int\tilde{J}\tilde{\phi}]\rho\right]\right]\\
 & \equiv & \exp[i\tilde{W}[\tilde{J}]],\nonumber 
\end{eqnarray}
where the tildes mean that the associated quantities are defined on
the closed time-contour: from $-\infty$ to $+\infty$ and than back
to $-\infty$ again. $T$ means the time ordering operation on this
contour, $J$ is an external source, and $\rho$ is the initial density
matrix for the field $\phi$. The trace operation is over the functions
on the closed time-contour. In the two by two matrix representation,
$\tilde{\phi}(x)=(\phi_{+}(x),\phi_{-}(x))$, $\tilde{J}[x]=(J_{+}(x),J_{-}(x))$,
and $\int\tilde{J}\tilde{\phi}=\int dxJ_{+}(x)\phi_{+}(x)-\int dxJ_{-}(x)\phi_{-}(x)$.
Note the extra minus sign in the above comes from the reversed time
contour part that has negative measure. A pair of variables $\phi_{\Delta}\equiv\phi_{+}(x)-\phi_{-}(x)$
and $\phi_{C}\equiv(\phi_{+}(x)+\phi_{-}(x))/2$ are also often used.
In the interaction picture: ${\cal L}[\phi]={\cal L}_{0}[\phi]-V[\phi]$,
we have,
\begin{equation}
\tilde{Z}[\tilde{J}]=\exp\left[-i\int V\left[\frac{1}{i}\frac{\delta}{\delta\tilde{J}}\right]\right]\exp[-\frac{i}{2}\int\int\tilde{J}(x)\tilde{G}_{0}(x,y)\tilde{J}(y)]{\rm Tr}(:\exp(i\int\tilde{J}\tilde{\phi}):\rho),
\end{equation}
where $\phi$ is in the interaction picture and $\tilde{G}_{0}$ is
a free propagator. We can develop perturbative calculations based
on the last expression. The C-number order parameter $\tilde{\varphi}$
is defined by 
\begin{equation}
\tilde{\varphi}(x)\equiv\frac{\delta\tilde{W}}{\delta\tilde{J}(x)}.
\end{equation}

Then the effective action $\tilde{\Gamma}$ is defined as the Legendre
transformation of $\tilde{W}$: 
\begin{equation}
\tilde{\Gamma}[\tilde{\varphi}]\equiv\tilde{W}[\tilde{J}]-\int\tilde{J}\tilde{\varphi}.
\end{equation}
The propagator part in the above $\tilde{J}(x)\tilde{G}_{0}(x,y)\tilde{J}(y)$
becomes 
\begin{equation}
J_{\Delta}(x)G_{R}(x,y)J_{C}(y)+J_{C}(x)G_{A}(x,y)J_{\Delta}(y)-iJ_{\Delta}(x)G_{C}(x,y)J_{\Delta}(y)\label{eq:2x2 propagator}
\end{equation}
where
\begin{eqnarray}
G_{R}(x,y) & = & i\theta\left(x^{0}-y^{0}\right)\left\langle \left[\phi\left(x\right),\:\phi\left(y\right)\right]\right\rangle ,\\
G_{A}(x,y) & = & -i\theta\left(y^{0}-x^{0}\right)\left\langle \left[\phi\left(x\right),\:\phi\left(y\right)\right]\right\rangle ,\\
G_{C}(x,y) & = & \left\langle \left\{ \phi\left(x\right),\:\phi\left(y\right)\right\} \right\rangle .
\end{eqnarray}
 The last term in Eq.(\ref{eq:2x2 propagator}) is special and imaginary.
It comes from the symmetric part of the propagator, while the rest
comes from the anti-symmetric part of the propagator. Thus they differ
by factor $i$. If used in the original equation, it yields the pure
Gaussian factor. Functionally Fourier transforming this term, we obtain
\begin{equation}
\exp[i\tilde{\Gamma}[\varphi_{\Delta,}\varphi_{C}]]=\int{\cal D}\xi P[\xi]\exp[i\tilde{S}_{eff}[\varphi_{\Delta,}\varphi_{C},\xi]],\label{eq:gen. eff. action}
\end{equation}
where 
\begin{equation}
P[\xi]=\exp[-\frac{1}{2}\int dx\int dy\xi(x)G_{C}(x,y)^{-1}\xi(y)],\label{eq: gaussian statistical weight}
\end{equation}
and 
\begin{eqnarray}
\tilde{S}_{eff}[\varphi_{\Delta,}\varphi_{C},\xi] & = & -\int dxdy\varphi_{\Delta}(x)G_{R}(x,y)^{-1}\varphi_{C}(y)-\int dxdy\varphi_{C}(x)G_{A}(x,y)^{-1}\varphi_{\Delta}(y)\label{eq: S_eff}\\
 &  & -\int dx\xi(x)\varphi_{\Delta}(x).\nonumber 
\end{eqnarray}
The full effective action in Eq.(\ref{eq:gen. eff. action}) is a
bundle of effective actions $\tilde{S}_{eff}[\tilde{\varphi},\xi]$
that depends on the field $\xi\left(x\right)$. This field can be
interpreted as the classical random field since the correlations of
them is generated by the Gaussian functional Eq.(\ref{eq: gaussian statistical weight}).
This function has a weight $P[\xi]$ in the average $\int{\cal D}\xi$\cite{Morikawa1986}.
The real part $\tilde{S}_{eff}[\tilde{\varphi},\xi]$ represents the
time evolution 
\begin{equation}
\frac{\delta\tilde{S}_{eff}[\tilde{\varphi},\xi]}{\varphi_{\Delta}(x)}|_{\varphi_{\Delta}=0}=-j_{C},
\end{equation}
which yields the equation of motion for $\varphi_{C}$: 
\begin{equation}
\int dy2G_{R}(x,y)^{-1}\varphi_{C}(y)+\xi=j_{C}.\label{eq:Langevin1}
\end{equation}
In this equation, the first term in the left hand side often yields
the friction term $\gamma\dot{\varphi}_{C}(x)+\gamma\dddot{\varphi}_{C}(x)+...$
originated from the time asymmetric part in the propagator. This time
reversal asymmetry comes from the choice of our initial condition
to choose in state, i.e. the closed time-contour from $-\infty$ to
$+\infty$ and than back to $-\infty$ again.

The correlation function for the random field $\xi$ is given by 
\begin{equation}
<...>_{\xi}=\int{\cal D}\xi...P[\xi],
\end{equation}
and 
\begin{equation}
\left\langle \xi(x)\xi(y)\right\rangle _{\xi}=G_{C}(x,y).
\end{equation}

The above separation of the full dynamics into the two parts, deterministic
$\tilde{S}_{eff}[\varphi_{\Delta,}\varphi_{C},\xi]$ and stochastic
$P[\xi]$, is general. The arguments are formal so far and actually
nothing special in equilibrium system. However in the non-equilibrium
settings, such as in the evolving background spacetime, the system
actually yields fluctuations and dissipation. 

Furthermore in our context, the IR divergence is only in the stochastic
part and the deterministic part is safe from the IR divergence. Therefore
the ordinary perturbation calculation is possible using $\tilde{S}_{eff}[\varphi_{\Delta,}\varphi_{C},\xi]$
and the stochastic part $P[\xi]$ agitates the system intermittently.
In the following sections we will see the detail of this structure.

\section{bi-linear interaction }

Let us consider the inflationary era in the early universe when the
cosmic expansion is exponential $a(t)=e^{Ht}=-(H\eta)^{-1}$, \textit{i.e.}
the de Sitter space-time
\begin{equation}
ds^{2}=dt^{2}-e^{2Ht}dx^{2}=(H\eta)^{-1}(d\eta^{2}-dx^{2}).
\end{equation}
This extreme exponential expansion is modeled to be caused by the
scalar field 
\begin{equation}
S[\phi]=\int d^{4}x(\partial_{\mu}\phi\partial^{\mu}\phi-m^{2}\phi^{2}-\xi R\phi)
\end{equation}
though only the massless minimally coupled case ($m=\xi=0$) is relevant.
The field is expanded in the normal mode on this space-time,

\begin{equation}
\phi(x)=\int\frac{d^{3}k}{(2\pi)^{3/2}}\frac{\sqrt{\pi}H\eta^{3/2}}{2}(\hat{a}_{\overrightarrow{k}}H_{\nu}^{(1)}(k\left|\eta\right|)e^{i\overrightarrow{k}\cdot\overrightarrow{x}}+\hat{a}_{\overrightarrow{k}}^{\dagger}H_{\nu}^{(2)}(k\left|\eta\right|)e^{-i\overrightarrow{k}\cdot\overrightarrow{x}}))
\end{equation}
where $\nu=(\frac{9}{4}-\frac{m^{2}\phi^{2}+\xi R\phi}{H^{2}})^{1/2}$
and $H_{\nu}^{(*)}$ are the Hankel functions. This normal mode is
selected by the requirement that the mode function reduces to the
Minkowski form locally $k\rightarrow\infty$ . We now restrict our
considerations to the most relevant massless minimally coupled case
$\nu=3/2$, 
\begin{equation}
H_{3/2}^{(1)}(k\left|\eta\right|)=-\frac{\sqrt{\frac{2}{\pi}}e^{ik\left|\eta\right|}(k\left|\eta\right|+i)}{(k\left|\eta\right|)^{3/2}}.
\end{equation}
The standard method is to introduce the separation of the field $\phi=\phi_{<}+\phi_{>}$at
around the scale of the horizon: $\phi_{>}\equiv\int dk\theta(k-H)\phi$\cite{starobinsky1982}.
And consider the interaction of them $\dot{\phi}_{>}(x)\dot{\phi}_{<}(x)$.
Then the effective dynamics for the large scale mode $\phi_{<}$is
given by integrating $\phi_{>}$ first. 

\begin{eqnarray}
\widetilde{Z}[\widetilde{J}] & = & \int{\cal D\widetilde{\phi}_{<}{\cal D\widetilde{\phi}_{>}}}\exp[i\widetilde{S}[\widetilde{\phi}]+i\int d^{4}x\widetilde{J}(x)\widetilde{\phi}(x)],\\
 & = & \int{\cal D\widetilde{\phi}_{<}}\exp[i\int d^{4}x\widetilde{\phi}_{<}(x)G_{0}(x-y)\widetilde{\phi}_{<}(y)+i\int d^{4}x\widetilde{J}(x)\widetilde{\phi}(x)],\nonumber \\
 & = & \int{\cal D\widetilde{\phi}_{<}}\exp[-\frac{1}{4}\int d^{3}k\phi_{<\Delta}(\overrightarrow{k})G_{C}(\overrightarrow{k})\phi_{<\Delta}(\overrightarrow{k})+\nonumber \\
 &  & +\int d^{3}k\phi_{<\Delta}(\overrightarrow{k})\theta(\Delta\eta)G_{R}(\overrightarrow{k})\phi_{<C}(\overrightarrow{k})+i\int d^{3}kJ(\overrightarrow{k})\phi(\overrightarrow{k})],\nonumber 
\end{eqnarray}
\begin{equation}
G_{0}=-\int\frac{d^{3}k}{(2\pi)^{3}}\frac{H^{2}}{2k^{3}}e^{-ik(\eta-\eta')+i\overrightarrow{k}\cdot(\overrightarrow{x}-\overrightarrow{x'})}(i-k\eta)(i+k\eta')
\end{equation}
\begin{eqnarray}
G_{C}(\overrightarrow{k}) & = & \frac{H^{2}}{k^{3}}\left((1+k^{2}\eta\eta')\cos(k\Delta\eta)+k\Delta\eta\sin(k\Delta\eta)\right)\propto\frac{H^{2}}{k^{3}}\\
G_{R}(\overrightarrow{k}) & =-i & \frac{H^{2}}{k^{3}}\left(-k\Delta\eta\cos(k\Delta\eta)+(1+k^{2}\eta\eta')\sin(k\Delta\eta)\right)\propto\frac{iH^{2}\Delta\eta}{k^{2}}\label{eq: de Sitter Gc and Gr}
\end{eqnarray}
Then in the last equation, the statistical and deterministic parts
are separated as 
\begin{eqnarray}
Z[J] & = & \int{\cal D\xi}P(\xi)\int{\cal D\phi_{<}}\exp[\int d^{3}k\phi_{<\Delta}(\overrightarrow{k})\theta(\Delta\eta)G_{R}(\overrightarrow{k})\phi_{<C}(\overrightarrow{k})\\
 &  & +i\int d^{3}kJ(\overrightarrow{k})\phi_{<}(\overrightarrow{k})+i\int d^{3}k\xi(\overrightarrow{k})\phi_{<}(\overrightarrow{k})],\nonumber 
\end{eqnarray}
where the statistical weight becomes the Gaussian form,
\begin{equation}
P(\xi)=\exp[-\frac{1}{4}\int d^{3}k\xi(\overrightarrow{k})G_{C}(\overrightarrow{k})^{-1}\xi(\overrightarrow{k})].
\end{equation}
The Langevin equation is derived by the variation by $\phi_{<}(\overrightarrow{k})$
to yield, 
\begin{equation}
3H\frac{d\phi(\overrightarrow{k})}{d\eta}=\xi
\end{equation}
and the correlation function of $\phi_{<}(\overrightarrow{k})$ becomes
\begin{equation}
\left\langle \phi_{<}(\overrightarrow{k})\phi_{<}(\overrightarrow{k})\right\rangle \thickapprox\frac{H^{2}}{k^{3}}
\end{equation}
at the Horizon crossing $\eta=-k^{-1}$, the standard evaluation point.
This is the stochastic method \cite{starobinsky1982,Hosoya1989}. 

However, artificial separation of free field $\phi=\phi_{<}+\phi_{>}$
at the Horizon does not resolve the quantum-classical transition problem.
The field $\phi_{<}$ is still quantum. Something equivalent to a
detector degrees of freedom is needed to discuss the statistical and
classical nature of the fluctuations in this formalism \cite{Nambu2009}.
We will further consider this point introducing the self interaction
of the scalar field in the effective action formalism in the next
section.

\section{self-coupled interaction}

In the above, we have no idea why the field $\phi_{<}$ behaves classically.
We would like to solve this problem together with the IR problem in
de Sitter space. We introduce the non-linearity of the scalar field
and the condensation of this quantum field. The action is given by
\begin{equation}
S[\phi]=\int d^{4}x(\partial_{\mu}\phi\partial^{\mu}\phi-m^{2}\phi^{2}-\xi R\phi-\lambda\phi^{4}/4!).
\end{equation}
The partition function becomes 
\begin{equation}
\widetilde{Z}[\tilde{J}]=\int{\cal D\tilde{\phi}}\exp[iS[\phi_{+}]-iS[\phi_{-}]+i\int d^{4}x\tilde{J}(x)\tilde{\phi}(x)]\equiv\exp i\widetilde{W},
\end{equation}
and its Legendre transform, i.e. the effective action becomes 
\begin{eqnarray}
\exp[i\widetilde{\Gamma}[\tilde{\varphi}]] & = & \exp i[\widetilde{W}[\widetilde{J}]-\int d^{4}x\widetilde{J}(x)\widetilde{\varphi}(x)]\\
 & = & \int{\cal D\widetilde{\phi}}\exp i[\widetilde{S}[\widetilde{\phi}]+\int d^{4}x\widetilde{J}(x)(\widetilde{\phi}(x)-\widetilde{\varphi}(x))],\nonumber \\
 & = & \int{\cal D\widetilde{\phi}}\exp i[\widetilde{S}[\widetilde{\varphi}+\widetilde{\phi}]+\int d^{4}x\widetilde{J}(x)\widetilde{\phi}(x)],\nonumber 
\end{eqnarray}
where the integration field is shifted by $\varphi$. Then expanding
the action around $\varphi$, we further have 
\begin{eqnarray}
\exp[i\widetilde{\Gamma}[\tilde{\varphi}]] & = & \exp i[W[\widetilde{J}]-\widetilde{J}\widetilde{\varphi}]\\
 & = & \int{\cal D\widetilde{\phi}}\exp i[\widetilde{S}_{int}[\widetilde{\phi};\widetilde{\varphi}]+\frac{1}{2}\int d^{4}x\widetilde{\phi}(x)G_{0}^{-1}(x-y)\widetilde{\phi}(y)-\int d^{4}x\widetilde{J}(x)\widetilde{\varphi}(x)],\nonumber \\
\nonumber 
\end{eqnarray}
where $\widetilde{S}_{int}[\phi;\varphi]$ is the Taylor expansion
of $\phi$ around $\varphi$. The first order term does not vanish
because we do not assume the$\varphi$ solves the free equation of
motion from $\widetilde{S}_{0}$ as in the ordinary stationary approach.
The second order term is absorbed into the propagator $G_{0}(x-y)$.
The third order or higher terms are genuine interactions which yields
the one particle irreducible graphs as usual. The first term yields
the factor in the effective action 

\begin{eqnarray}
 &  & \exp[i\widetilde{S}_{0}[\tilde{\varphi}]]\int{\cal D\widetilde{\phi}}\exp i[\lambda\varphi^{3}\phi]+\frac{1}{2}\int d^{4}x\phi(x)G_{0}^{-1}(x-y)\phi(y)]\\
 & = & \exp[i\widetilde{S}_{0}[\tilde{\varphi}]]\int{\cal D\widetilde{\phi}}\exp i[(\lambda\varphi(x)^{3})_{\Delta}G_{R}(x-y)(\lambda\varphi(y)^{3})_{C}+(\lambda\varphi(x)^{3})_{C}G_{A}(x-y)(\lambda\varphi(y)^{3})_{\Delta}\nonumber \\
 &  & +i(\lambda\varphi(x)^{3})_{\Delta}G_{C}(x-y)(\lambda\varphi(y)^{3})_{\Delta}]\nonumber 
\end{eqnarray}
where $G_{R}(x-y),\:G_{C}(x-y)$ have the form Eq.(\ref{eq: de Sitter Gc and Gr}).
In the interaction, the remaining kinetic terms becomes irrelevant
in the IR limit and the mass term does not exist. The IR divergent
term $G_{C}(x-y)$ becomes pure Gaussian and therefore can be separated
as in the previous way to yield statistical fluctuations. The finite
terms $G_{R}(x-y),\:G_{A}(x-y)$ yield friction term. However the
macroscopic friction term $-3H\dot{\varphi}$ directly associated
with the cosmic expansion in the equation of motion dominates this
friction. Thus the full effective action terns out to be
\begin{equation}
\exp[i\widetilde{\Gamma}[\tilde{\varphi}]]=\int{\cal D\xi}P(\xi)\exp[i\Gamma[\tilde{\varphi;\xi}]],
\end{equation}
 where

\begin{eqnarray}
\exp[i\Gamma[\tilde{\varphi;\xi}]] & = & \exp[i\widetilde{S}_{0}[\tilde{\varphi}]]\exp i[S'_{int}[\frac{\delta}{i\delta\widetilde{J}};\varphi]]\exp i[\frac{1}{2}\int d^{4}x\widetilde{J}(x)G'_{0}(x-y)\widetilde{J}(y)\\
 &  & +\int d^{4}x\xi(x)(x)(\lambda\varphi(x)^{3})_{\Delta}].\nonumber 
\end{eqnarray}
where $S'_{int}$is the interaction term with the linear term removed
and $G'_{0}(x-y)$ is the propagator with the IR divergence removed. 

This allows the ordinary perturbation calculations, infrared safe,
for higher order quantum corrections; the infrared diverging term
is fully separated in the fluctuation kernel $P(\xi).$ The statistical
fluctuations represented by $\xi$ acts on the local quantum dynamics
intermittently. However the effect is mostly limited in the long range
IR region. 

We can obtain the equation of motion for the order parameter $\tilde{\varphi}$:
\begin{equation}
\frac{\delta\tilde{\Gamma}}{\delta\tilde{\varphi}(x)}=-\tilde{J}(x).
\end{equation}
This becomes in the lowest order of $\xi_{k}$ in the strong damping
regime, 
\begin{equation}
3H\dot{\varphi}{}_{k}+(\lambda/2)\varphi_{0}^{2}\varphi_{k}=(\lambda/2)\varphi_{0}^{2}\xi_{k}.
\end{equation}
This equation yields the same power spectrum for $\varphi_{k}$ but
a slightly different amplitude: 
\begin{equation}
\left\langle \varphi_{k}\varphi_{k}\right\rangle _{\xi}\thickapprox\lambda^{2}\varphi_{0}^{4}\frac{H^{2}}{k^{3}}.
\end{equation}
The spectrum does not change because the statistical fluctuations
showing IR divergence dominate in the full quantum propagators. The
amplitude does change because the statistical fluctuations are extracted
through the non-linear interactions. 

There are variety of applications of the obtained Langevin equation
to the actual inflationary dynamics. This can be used to select the
correct model among fair amount of inflationary models presently proposed.
Interestingly, the Langevin analysis on the original standard model
of inflation yields the same result as the bi-linear case. The detail
will be reported elsewhere.

\section{Non-minimal massive case}

The massless minimal coupling scalar field is most useful for inflation.
Therefore the IR divergence has been considered first in this case.
However the IR anomalous enhancement is not restricted to this case.
We will see the general scalar field in de Sitter spacetime now. The
same approach is enough for this purpose. 

The propagator becomes \cite{Bunch1978} 
\begin{equation}
\left\langle \phi(x)\phi(x')\right\rangle =\int\frac{d^{3}k}{(2\pi)^{3}}H_{\nu}^{(1)}(k\eta)H_{\nu}^{(2)}(k\eta'),
\end{equation}
where the real and imaginary part of the Hankel functions are manifest,
\begin{equation}
H_{\nu}^{(1)}(z)=J_{\nu}(z)+iY_{\nu}(z),\:H_{\nu}^{(2)}(z)=J_{\nu}(z)-iY_{\nu}(z),
\end{equation}
and behaves, in small argument, as 
\begin{equation}
J_{\nu}(z)=\frac{2}{\Gamma(1+\nu)}z^{\nu}+O(z^{2+\nu}),\:Y_{\nu}(z)=-\frac{2^{\nu}\Gamma(\nu)}{\pi}z^{-\nu}+O(z^{2-\nu}).
\end{equation}
Therefore the IR behavior is obvious: 
\begin{eqnarray}
G_{C}(\overrightarrow{k}) & = & J_{\nu}(k\eta)J_{\nu}(k\eta')+Y_{\nu}(k\eta)Y_{\nu}(k\eta')\propto H^{2\nu}k^{-2\nu},\\
G_{R}(\overrightarrow{k}) & =i( & J_{\nu}(k\eta)Y_{\nu}(k\eta')-J_{\nu}(k\eta')Y_{\nu}(k\eta))\propto k^{0}.\nonumber 
\end{eqnarray}
Since $\nu=(\frac{9}{4}-\frac{m^{2}\phi^{2}+\xi R\phi}{H^{2}})^{1/2}$
, and therefore $0\leqq\nu\leqq3/2$, it is apparent that the IR dangerous
term exists only in $G_{C}(\overrightarrow{k})$. This term is isolated
from the microscopic dynamics as a statistical fluctuations as before.
In the present general case, the IR behavior is milder than the massless
minimal case. 

However higher loop contributions and/or higher point functions may
yield severe IR behavior\cite{Polyakov2008}. Even in those cases,
it may happen that the IR divergence is associated with the imaginary
part of the effective action. For example, the graph is associated
with the real particle emission process. Then these IR divergent terms
can be separated from the quantum dynamics as the statistical weight.
In this case the statistical weight is no longer the Gaussian form.
Therefore unusual statistical mechanics is expected. We don't know
how extent various IR divergence can be absorbed into the imaginary
part of the effective action at present. We hope we can report this
interesting problem soon in our future publications.

\section{conclusions and prospects}

We studied the non-equilibrium aspect of the inflationary phase transition
in the early Universe. 

The massless minimally couple scalar field yields the exponential
expansion of the Universe and the quantum fields on this spacetime
becomes peculiar and yields IR divergence. This IR divergence is quite
relevant to produce the seed fluctuations of all the structures in
the Universe. On the other hand this IR divergence destroys the quantum
field theory and the perturbation method. 

In this paper, we clarified that this dilemma comes from the mixing
up the finite quantum part and the diverging statistical part in the
formalism. By separating these two contributions, we could derive
the classical Langevin equation of motion for the order parameter
of the inflationary phase transition. This IR divergence simply reflects
that the statistical fluctuations have long-time correlation. This
Langevin equation is the manifestly classical evolution and is adequate
to describe the macroscopic dynamics such as the large scale structures
in the Universe. On the other hand the remaining quantum evolution
is free from IR divergence. This separation of statistical and quantum
fluctuations has been the crucial point of the problem. 

We further need to clarify the problem why and how extent the IR divergence
is associated with the imaginary part or the statistical part of the
effective action. For the scalar fields in de Sitter space, this association
was general. Probably the IR divergence and the related peculiar non-equilibrium
behavior in the curved space comes from the violent particle production
process from the vacuum. Applying the Bogoliubov transformation formalism,
we would like to generalize the present work to other evolving space-times. 
\begin{acknowledgments}
The author thanks to Gi-Chol Cho, Takeshi Fukuyama, Kyuzo Teshima
and Akio Sugamoto for fruitful discussions. He also thanks to all
the astrophysics laboratory members at Ochanomizu university for inspiring
discussions and criticisms. \end{acknowledgments}


\begin{thebibliography}{10}
\bibitem{baumann2009}D. Baumann, http://arxiv.org/abs/0907.5424. 

\bibitem{Guth1981}A. H. Guth, Phys. Rev. D 23, 347 (1981).

\bibitem{sato1981} K. Sato, Monthly Notices of Royal Astronomical
Society, 195, 467, (1981).

\bibitem{arai2012}T. Arai, Phys. Rev. D86 104064 (2012). 

\bibitem{Polyakov2008}A. M. Polyakov, Nucl. Phys. B, Proc. Suppl.
B797, 199 (2008).

\bibitem{Morikawa1987}M. Morikawa, Prog. Theor. Phys. \textbf{77},
1163 (1987).

\bibitem{Nambu2009}Y. Nambu and Y. Ohsumi, PRD\textbf{80},124031
(2009); PRD\textbf{84}, 044028 (2011).

\bibitem{Fukuyama2009} T. Fukuyama and M. Morikawa, Phys. Rev.\textbf{
D80}, 063520 (2009).

\bibitem{Morikawa1986}M. Morikawa, Phys. Rev. D\textbf{33} (1986),
3607. 

\bibitem{starobinsky1982}Starobinsky, Alexei A. (1982). Phys. Lett.
B117: 175\textendash 8.

\bibitem{keldish1964}Keldysh, L. V., 1964, Zh. Eksp. Teor. Fiz.,
47, 1515; {[}Sov. Phys. JETP, 1965, 20, 1018{]}. 

\bibitem{kadanoff1962}Kadanoff, L. P., and Baym, G., 1962, Quantum
Statistical Mechanics, (Benjamin, New York).

\bibitem{Hosoya1989}A. Hosoya, M. Morikawa, and K. Nakayama, Int.
J. Mod. Phys. \textbf{A 04}, 2613 (1989). 

\bibitem{Bunch1978}T. S. Bunch and P. C. W. Davies, Proc. R. Soc.
Lond. A 360, 117 (1978).\end{thebibliography}
\end{document}